\documentclass[letterpaper]{article}
\usepackage{spconf,amsmath,graphicx}
\usepackage{subfigure}
\usepackage{amssymb, mathrsfs}
\usepackage{cite}

\newtheorem{theorem}{Theorem}[section]

\newtheorem{proposition}[theorem]{Proposition}

\title{Compressed Sensing Recoverability In Imaging Modalities}
\name{Mahdi S. Hosseini and Konstantinos N. Plataniotis}%\thanks{Thanks to XYZ agency for funding.}}
\address{ECE Department, University of Toronto}

\begin{document}
%\ninept
%
\maketitle
\begin{abstract}
The paper introduces a framework for the recoverability analysis in compressive sensing for imaging applications such as CI cameras, rapid MRI and coded apertures. This is done using the fact that the Spherical Section Property (SSP) of a sensing matrix provides a lower bound for unique sparse recovery condition. The lower bound is evaluated for different sampling paradigms adopted from the aforementioned imaging modalities. In particular, a platform is provided to analyze the well-posedness of sub-sampling patterns commonly used in practical scenarios. The effectiveness of the various designed patterns for sparse image recovery is studied through numerical experiments.
\end{abstract}
\section{Introduction}
\label{sec:intro}
Recoverability in Compressed Sensing (CS), in other words the problem of how to effectively reconstruct a sparse signal from its compressive measurements, remains an open research problem for both engineers and applied mathematicians \cite{CandesTao:2005, KashinTemlyakov:2007, Zhang:2008, BrucksteinDonohoElad:2009}. The demand for such analysis is paramount in CS imaging modalities where there is a need to examine the effectiveness of structured sensing architectures subject to hardware configuration constraints \cite{CandesRombergTao1:2006, DuarteEldar:2011}. Examples include, but not limited to, Compressed Imaging (CI) cameras \cite{DuarteDavenportTakharLaskaTingKellyBaraniuk:2008, OikeGamal:2013}, rapid Magnetic Resonance Imaging (MRI) for medical applications \cite{LustigDonohoPauly:2007, CandesRombergTao1:2006} and coded apertures in optics \cite{MarciaWillett:2008}.

To date, there are only a few limited studies examining the conditions needed for unique recovery of sparse signals in CS imaging modalities. These works extend the results obtained for one dimensional CS recoveries to Kronecker sampling structures \cite{Jokar:2010, DuarteBaraniuk:2011, DuarteEldar:2011, CaiafaCichocki:2012} by utilizing mutual coherency and restricted isometry analysis of the sensing matrices \cite{DonohoElad:2003, CandesTao:2005}. Unfortunately, such analysis constitute an NP-hard problem and provides loose recovery bounds suitable only for specialized sensing modalities such as the Gaussian/Bernoulli matrices \cite{KashinTemlyakov:2007, Zhang:2008, BrucksteinDonohoElad:2009}. Recent analysis results introduced in \cite{AspremontGhaoui:2011, Aspremont:2011, HosseiniFazeliPlataniotis:2012, LeeBresler:2008, TangNehorai:2011} provide a new insight for tracking the needed recovery bounds using relaxed Semidefinite Programming (SDP) methods. Following these developments, this paper studies the recoverability conditions for sparse signals in CS imaging modalities that are characterized in terms of a (given):\vspace{-.1in}

\begin{itemize}
\item \textit{Sub-Sampling pattern:} which defines the under-sampling process in the domain of its feasibility. For example, in tomography imaging, CS samples are taken via projection along the radial lines of beam-paths and is modelled in a 2-D Fourier mask \cite{CandesRombergTao1:2006}. \vspace{-.1in}
\item \textit{Sensing modality:} which defines the connection between the acquisition and sampling domains, thus identifying the sampling basis for CS recovery, e.g. Fourier basis in MRI \cite{LustigDonohoPauly:2007}.\vspace{-.1in}
\end{itemize}

In this paper we address the recoverability of three different CS imaging architectures i.e. CI cameras, rapid MRI and coded apertures. In particular, we study the effectiveness of different sensing modality for sampling with their pertinent sub-sampling pattern masks. The unique sparse recovery conditions of such modalities are analyzed by tracking the lower bound of Spherical Section Property (SSP) of the corresponding sensing matrices with Kronecker structures. This lower bound is calculated using a relaxed Semidefinite Programming (SDP) method in \cite{HosseiniFazeliPlataniotis:2012}. The bound provides a platform to compare the well-posedness of any designed CS architecture prior to its practical developments, which can be an expensive and time consuming task. Furthermore, we have tested the sparse recovery of the constructed sensing matrices for the aforementioned applications by means of  $\ell_1$-minimization using Basis Pursuit (BP) problem in \cite{BergFriedlander:2008} to validate the recovery bounds provided by the relaxed-SDP.

\section{General CS Imaging Pipeline}\label{Sec:CSImaging}
\begin{figure}[h]
  \centerline{\includegraphics[width=8.5cm]{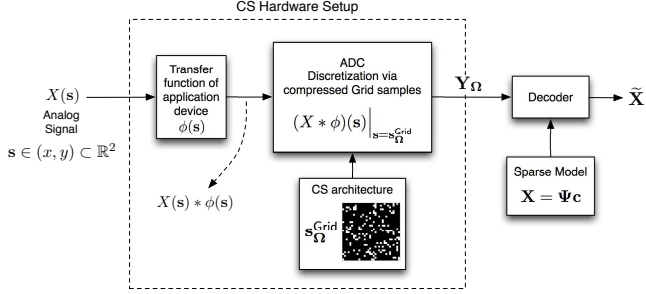}}
  \caption{CS pipline for image acquistion and reconstruction}
  \label{CSpipeline}
\end{figure}

Following the works done in \cite{DuarteDavenportTakharLaskaTingKellyBaraniuk:2008, OikeGamal:2013, LustigDonohoPauly:2007, MarciaWillett:2008, HahnLimChoiHorisakiBrady:2011, ReddyVeeraraghavanChellappa:2011}, the common framework for existing CS imaging applications can be generally overviewed by the pipeline demosntrated in Figure \ref{CSpipeline}. The rational behind such generalization is to modify the recoverability of the existing CS imaging applications under one unique analysis framework in order to study the impact of different sampling scenarios. In this pipeline, the signal (image) of interest $X({\bf s})$ is realized in a domain of a linear transfer function, i.e. $X(\bf{s})\ast\phi(\bf{s})$, which is defined based on the characteristics of the applied system. For instance, in MRI \cite{LustigDonohoPauly:2007}, the image is observed in Fourier transform domain (also known as k-space), where it holds the raw data representing  the spin density distributions. Therefore, the corresponding transfer function is defined by the Fourier kernels i.e. $\phi(\bf{s})= \mathscr{F}(\bf{s})$.

The maximum allowable sensory resolution is carried out by the mesh design of the Field-Of-View (FOV) in the application. For example, in CI camera \cite{DuarteDavenportTakharLaskaTingKellyBaraniuk:2008, OikeGamal:2013}, the number of the grid sampling points in the image is defined by the number of the micro-mirror arrays deployed in the FOV. Such grids are modelled by ${\bf s}^{\text{Grid}}\in\mathbb{R}^m\times\mathbb{R}^n$ and provide an $m$-by-$n$ discrete image \cite{OikeGamal:2013, BourquardAguetUnser:2010}. Hence, the full discrete sampled image is provided by $Y = (X\ast\phi)({\bf s}){\Big{\vert}}_{{\bf s}^{\text{Grid}}}$ and this system is known to be fully determined for any mathematical purposes. In general, the acquired samples are discrete and they can considered to be defined by the following discretization, i.e.
\begin{equation}\label{PF1}
Y(x,y) = \sum^{m-1}_{i=0}\sum^{n-1}_{j=0}{X[i,j]\phi(\frac{y}{\Delta_1}-i)\phi(\frac{x}{\Delta_2}-j)},
\end{equation}
where, $\Delta_1$ and $\Delta_2$ are the spatial grid spacing for vertical and horizontal vertices, accordingly and $X[i,j]$ is the discrete model of the original image space.

\begin{figure}[h]
\centerline{
\subfigure[]{\includegraphics[width=0.09\textwidth]{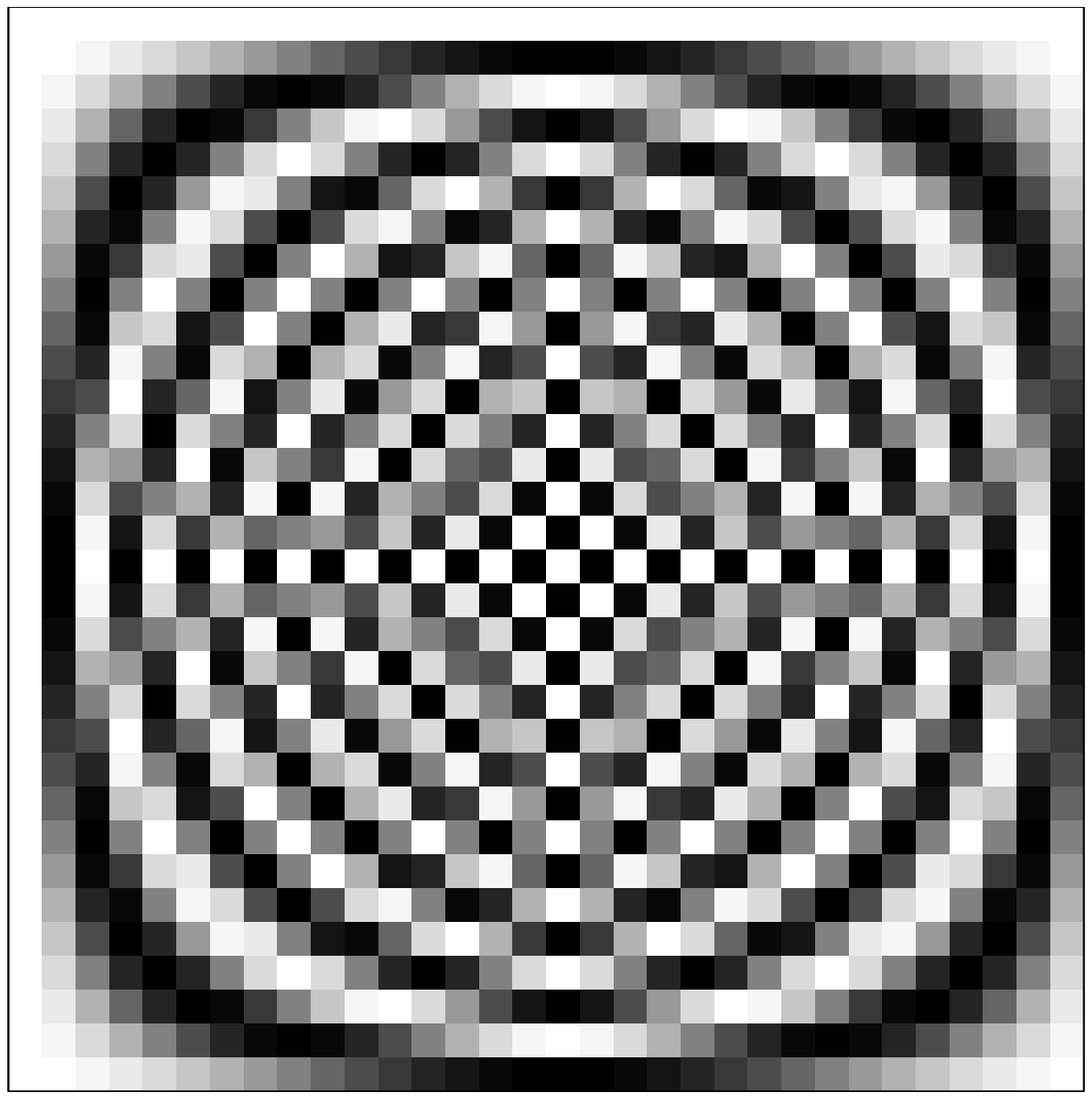}\label{fig:RFourier}}
\subfigure[]{\includegraphics[width=0.09\textwidth]{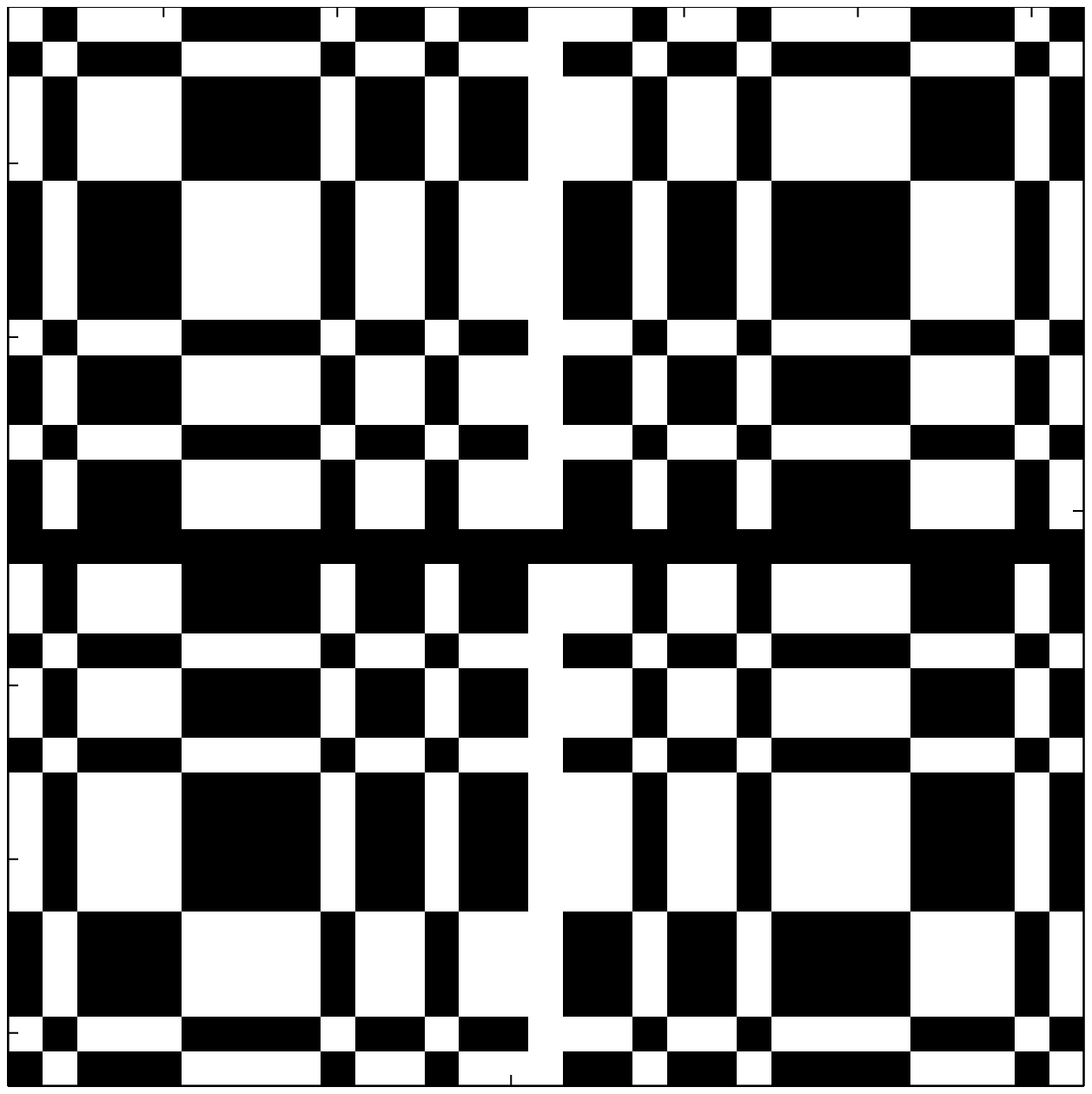}\label{fig:MURA}}
\subfigure[]{\includegraphics[width=0.09\textwidth]{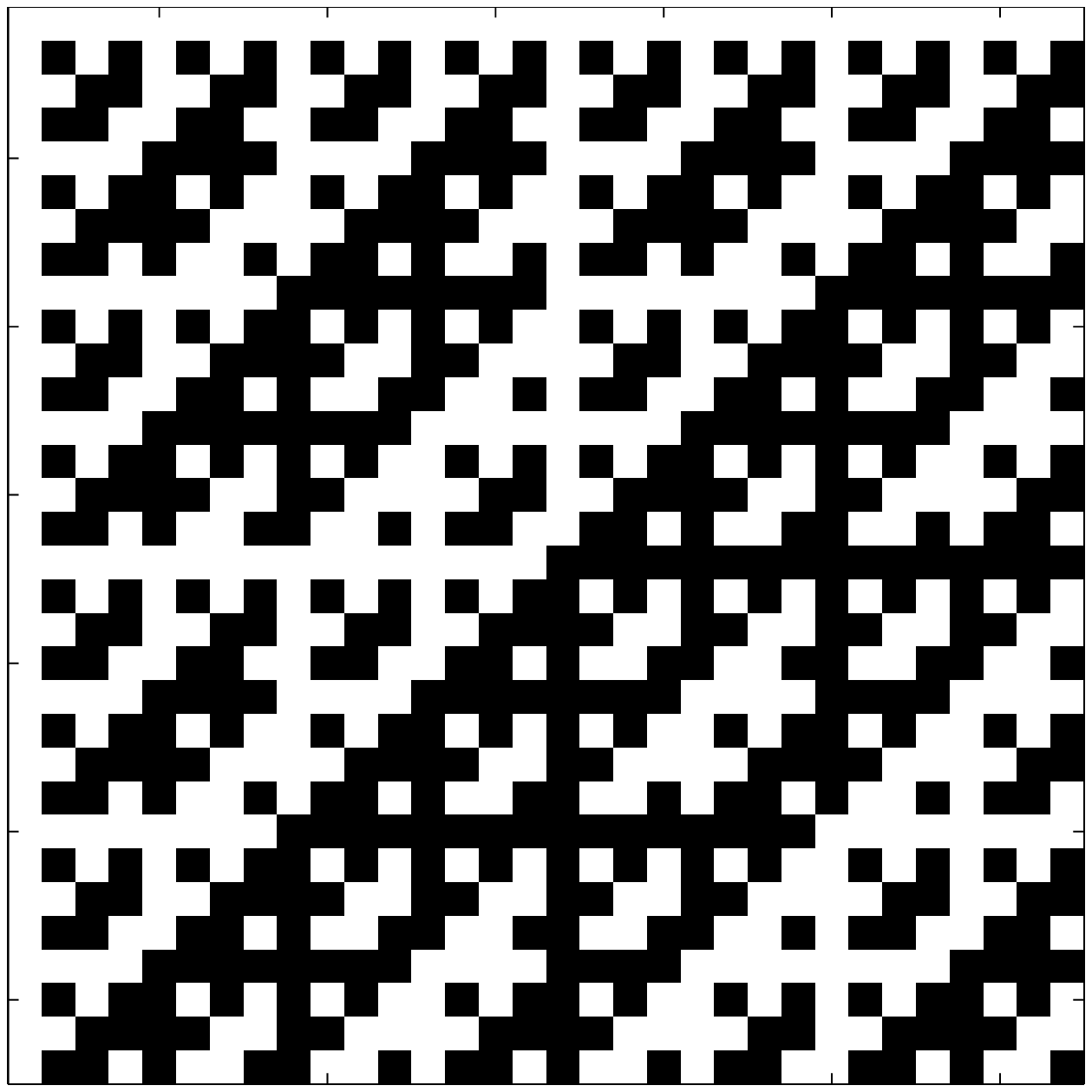}\label{fig:Walsh}}
\subfigure[]{\includegraphics[width=0.09\textwidth]{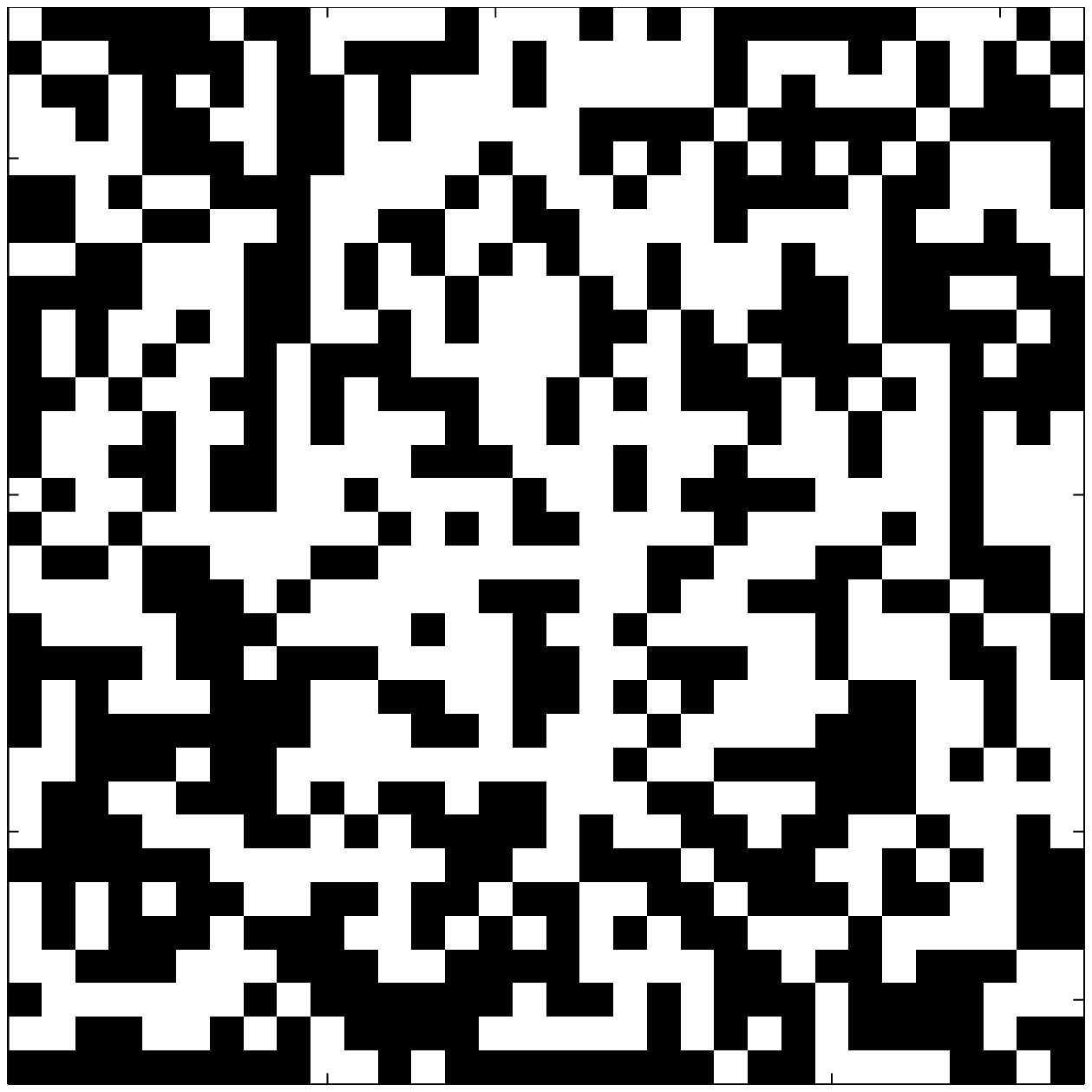}\label{fig:Bernoulli}}
\subfigure[]{\includegraphics[width=0.09\textwidth]{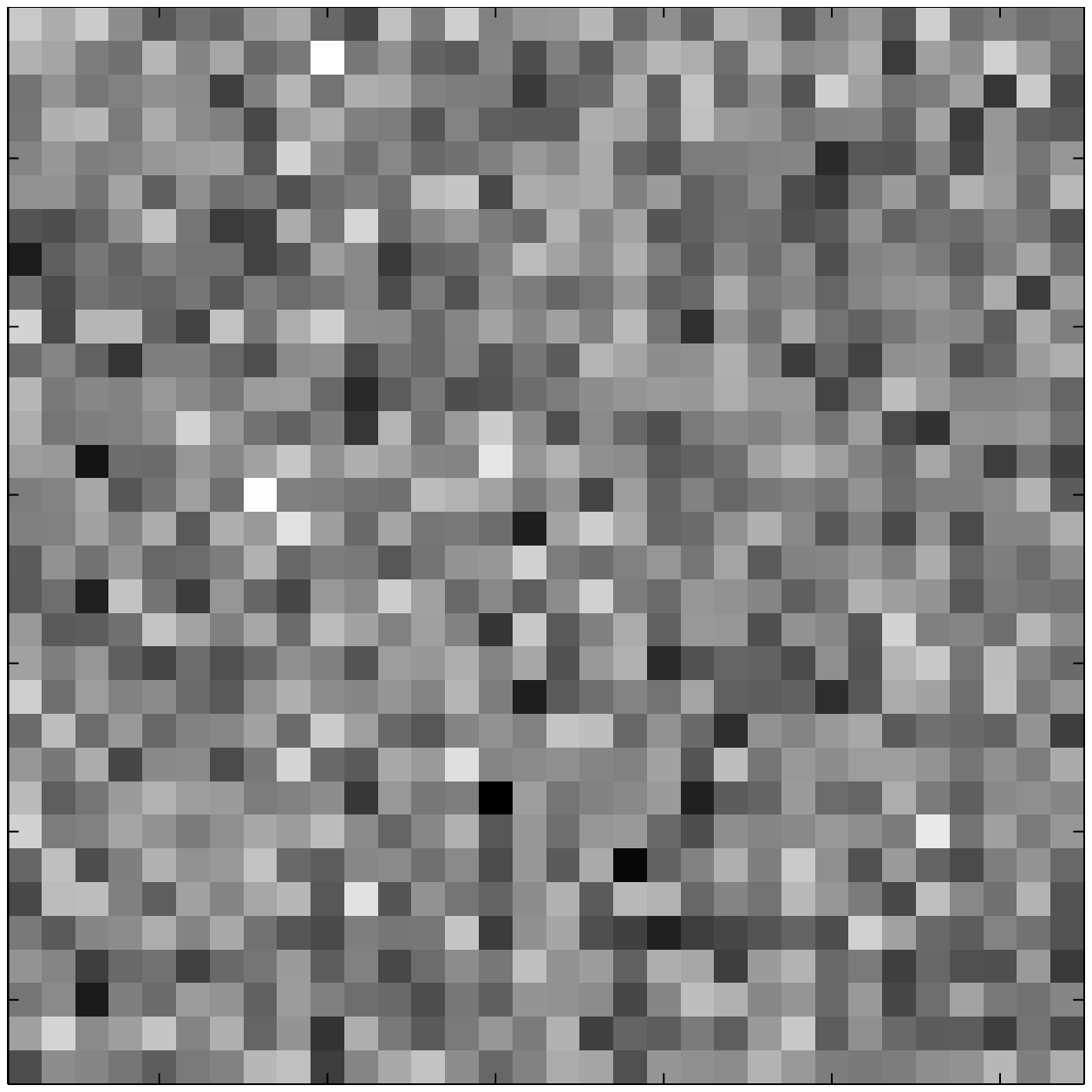}\label{fig:Gaussian}}
}\vspace{-.15in}
\caption{Sampling bases $\Phi$: (a) Fourier; (b) MURA pattern; (c) Walsh-Hadamard; (d) Bernoulli; (e) Gaussian}
\label{Fig:SamplingBases}
\end{figure}

In CS imaging application, a subset of discrete samples are available for digitization i.e. ${\bf s}^{\text{Grid}}_{\Omega}$, where $\Omega\subset\{0,1,\hdots,m\cdot n-1\}$ is a subset indices defined from the mesh design. Therefore, the compressed samples, practically, are given by $Y_{\Omega} = (X\ast\phi)({\bf s}){\Big{\vert}}_{{\bf s}^{\text{Grid}}_{\Omega}}$ and mathematically is known to be underdetermined system \cite{CandesRombergTao1:2006, KashinTemlyakov:2007, BrucksteinDonohoElad:2009}. This is because the number of the acquired samples are less than the total number of the available grid points i.e. $|\Omega|<m\cdot n$. Later on, the digitized values of the CS samples are carried out by an Analog-to-Digital (ADC) converter of the CS hardware and can be represented by ${\bf Y} = {\Phi}^T_1 {\bf X}  {\Phi}_2$, where, ${\bf X}$ and ${\bf Y}$ are the original and transferred images in digital format, respectively. Also, ${\Phi}_1\in\mathbb{R}^{m\times m}$ and ${\Phi}_2\in\mathbb{R}^{n\times n}$ are the discrete basis dictionaries exemplified in Figure \ref{Fig:SamplingBases} for different sampling (feasibility) domain. The corresponding Kronecker tensor arraying can be driven by $vec({\bf Y}) = {\bf\Phi}^T\cdot vec({\bf X}) = (\Phi_2\otimes\Phi_1)^T vec({\bf X})$, where $vec(\cdot)$ denotes mode-$1$ vectors in one column (lexicographic ordering \cite{Brewer:1978}. Now, we can subject the assumption of compressed sampling to the system of measurements by
\begin{equation}\label{PF3}
vec({\bf Y}_{\Omega})= {\bf\Phi}^T_{\Omega} vec({\bf X}),
\end{equation}
where, ${\bf\Phi}_{\Omega}$ are the subset of column bases of $\bf\Phi$ subjected to the compressed sampling indices $\Omega$.
As mentioned, the measurement system in (\ref{PF3}) is underdetermined and, therefore, there are infinite possible solutions to fit this equation. Nevertheless, the prior information from the original signal makes the decoding possible by means of sparse decomposition in a pre-defined basis domain i.e. ${\bf X}=\Psi{\bf C}\Psi^T $ and by substituting it in (\ref{PF3}) yields
\begin{equation}\label{PF4}
vec({\bf Y}_{\Omega})= {\bf\Phi}^T_{\Omega}{\bf\Psi} vec({\bf C}),
\end{equation}
where ${\bf\Psi}=\Psi\otimes\Psi$ is the Kronecker representation of the sparse operator. We define the sensing matrix $A={\bf\Phi}^T_{\Omega}{\bf\Psi}\in\mathbb{R}^{|\Omega|\times{N}}$, where $N=m\cdot n$ is the number of the image pixels.

The sub-sampling pattern $\Omega$ is usually dependent to the architecture of the CS hardware. We adopt four masks from literature and demonstrate in Figure \ref{tab:CS_Application} for CS imaging applications: radial \cite{CandesRombergTao1:2006}, random \cite{DuarteDavenportTakharLaskaTingKellyBaraniuk:2008, OikeGamal:2013,LustigDonohoPauly:2007}, density-varied \cite{LustigDonohoPauly:2007}, and down-sampling \cite{MarciaWillett:2008}. Table \ref{tab:CS_Application} lists three main applications, where such masks are used for under-sampling. \vspace{-.05in}
\begin{figure}[h]
\centerline{
\subfigure[]{\includegraphics[width=0.09\textwidth]{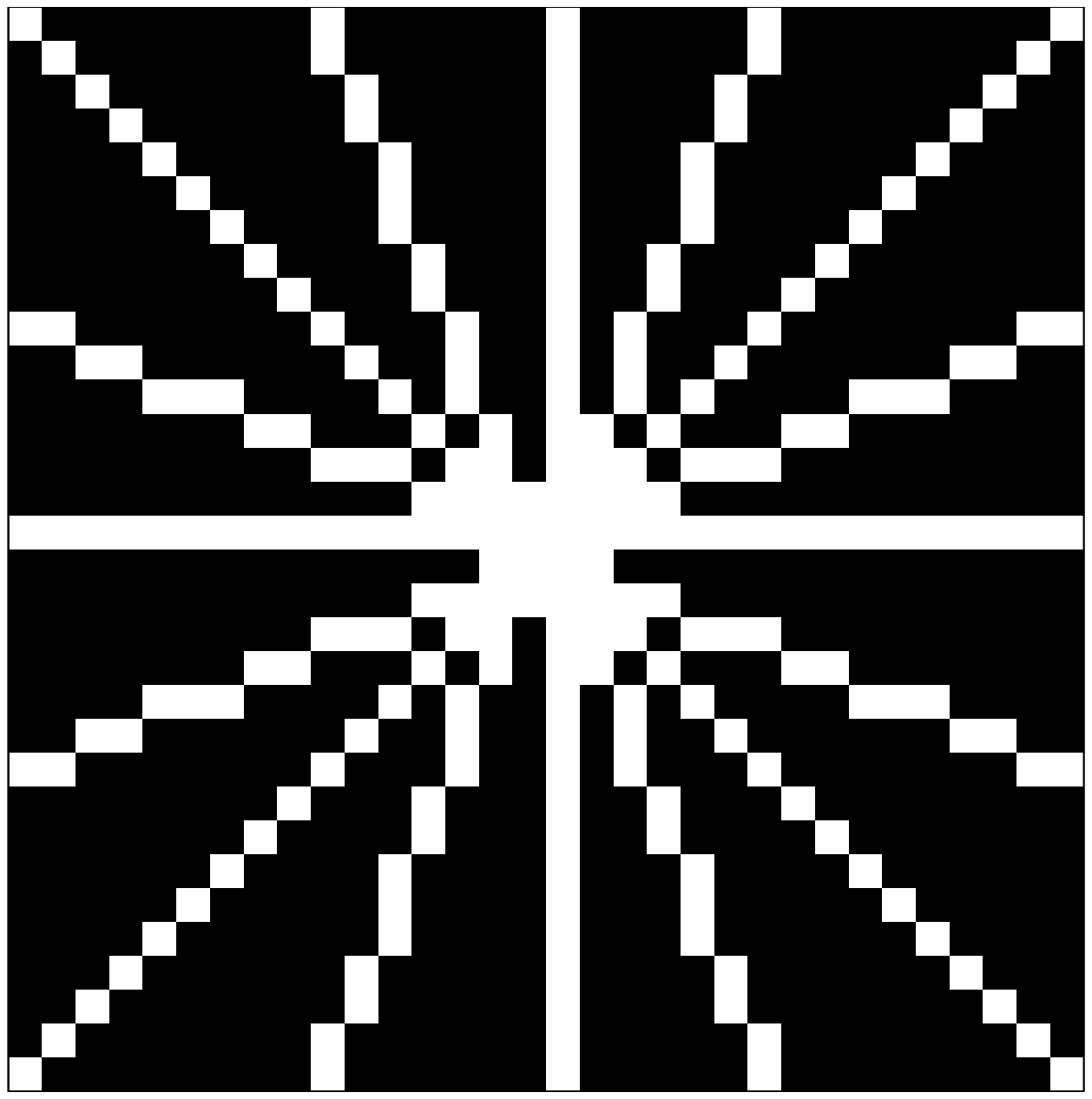}
\label{Fig:RadialMask}}
\subfigure[]{\includegraphics[width=0.09\textwidth]{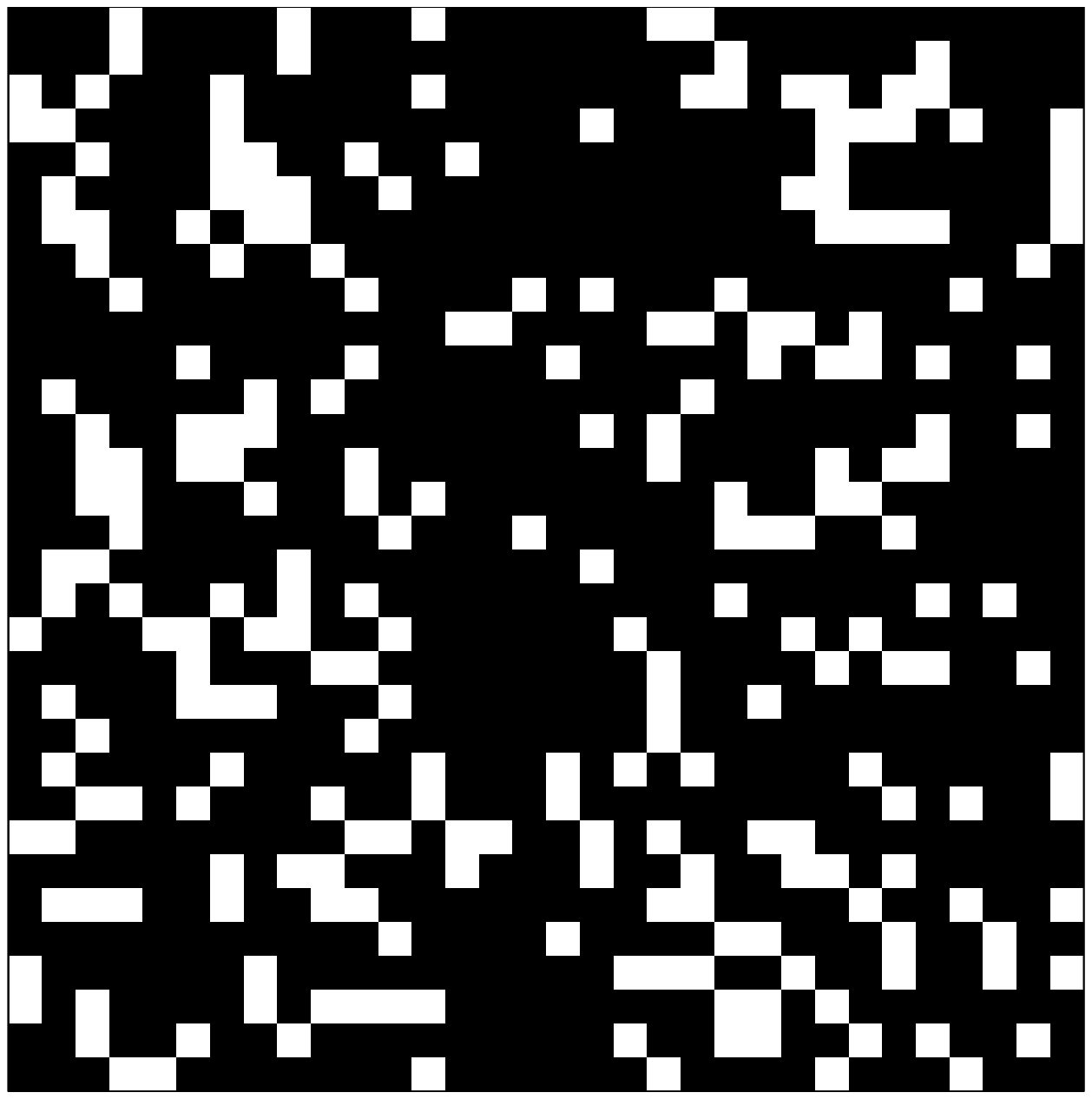}
\label{Fig:BernoulliMask}}
\subfigure[]{\includegraphics[width=0.09\textwidth]{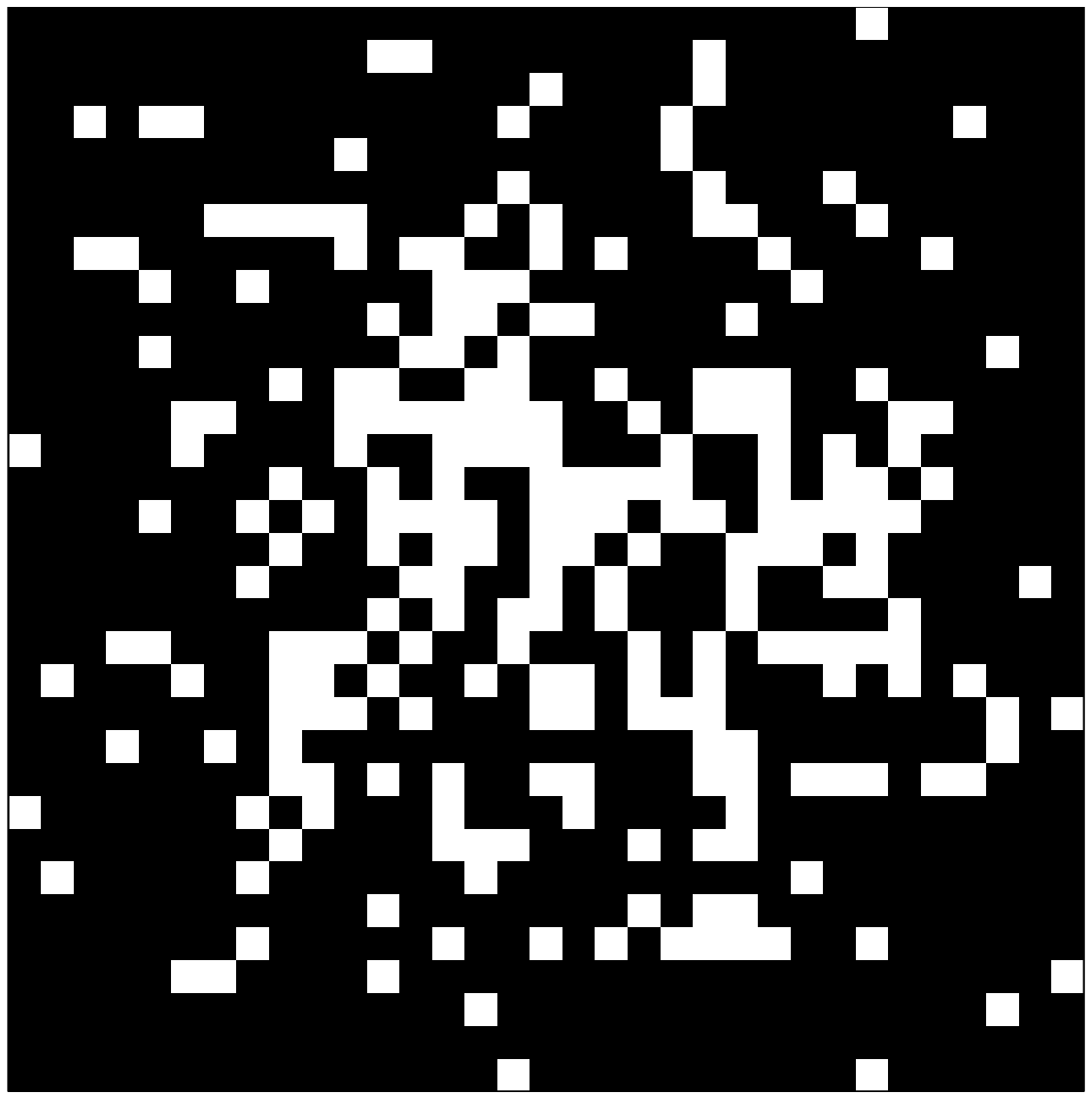}
\label{Fig:WeightedMask}}
\subfigure[]{\includegraphics[width=0.09\textwidth]{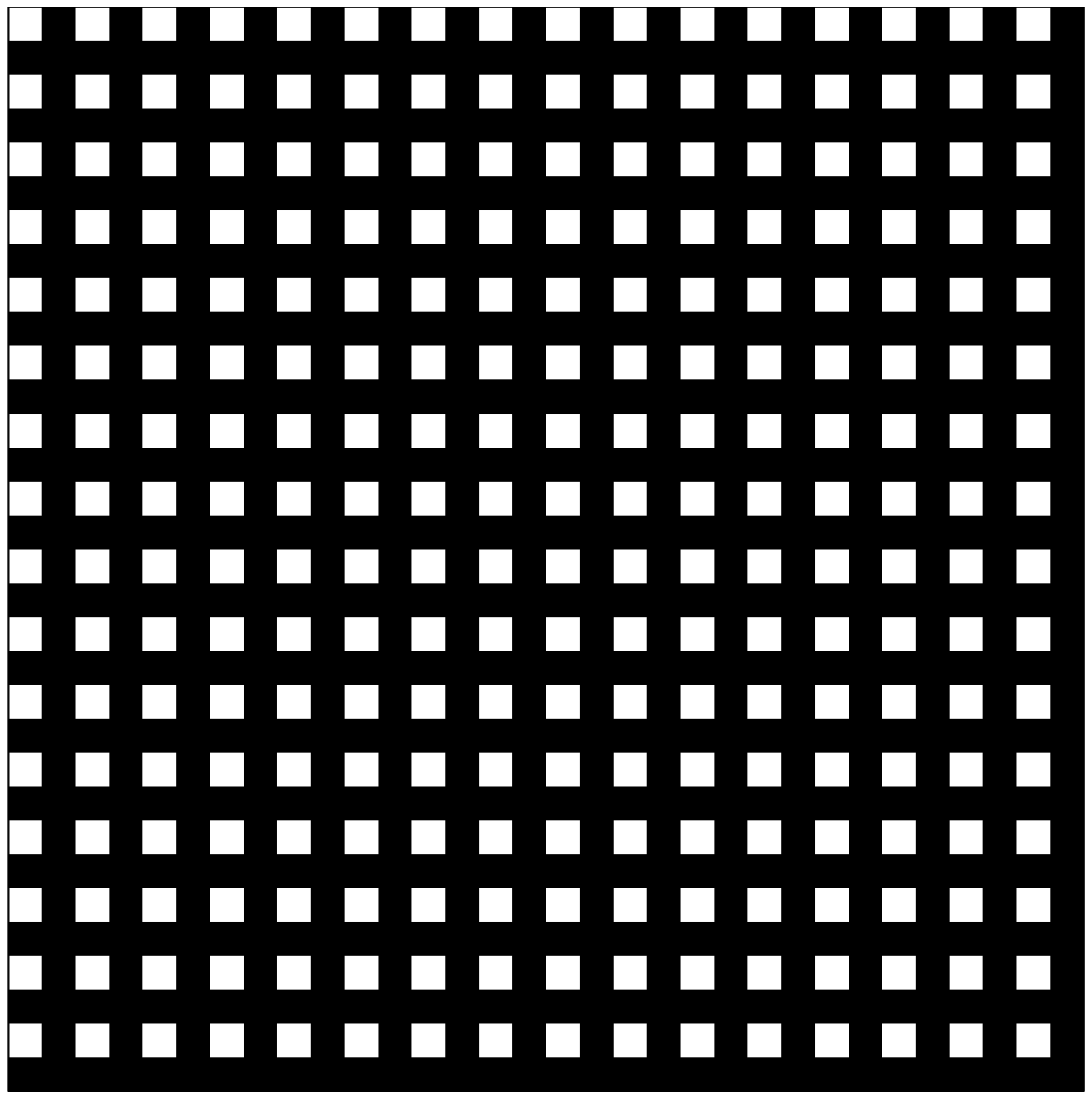}
\label{Fig:ApertureMask}}
}\vspace{-.15in}
\caption{Four sub-sampling $32\times 32$ CS masks i.e. ${\bf s}^{\text{Grid}}_\Omega$: (a) $8$ radial lined mask; (b) random (uniform distribution); (c) density-varied; (d) down-sampling by factor $4$.}
\label{fig:Masks}
\end{figure}\vspace{-.2in}
\vspace{-.3in}
\begin{table}[h]
\caption{CS Imaging application charactristics} % title name of the table
\centering % centering table
\begin{tabular}{p{.14\textwidth} p{.1\textwidth} p{.08\textwidth} cp{.04\textwidth}} % creating 10 columns
\hline\hline% inserting double-line
{\bf CS Application} & $\Phi$ & {\bf CS Mask} & {\bf Ref.}\\ \hline\hline % inserts single-line
CI Camera & Fig.\ref{Fig:SamplingBases}(c-e) & Fig.\ref{Fig:BernoulliMask} & \cite{DuarteDavenportTakharLaskaTingKellyBaraniuk:2008, OikeGamal:2013}\\ \hline
Rapid MRI & Fig.\ref{Fig:SamplingBases}(a)  & Fig.\ref{fig:Masks}(a-c) & \cite{CandesRombergTao1:2006,LustigDonohoPauly:2007} \\ \hline
Coded Apertures & Fig.\ref{Fig:SamplingBases}(b) & Fig.\ref{Fig:ApertureMask} & \cite{MarciaWillett:2008}\\[1ex]
\hline % inserts single-line
\end{tabular}
\label{tab:CS_Application}
\end{table}\vspace{-.2in}

\section{Unique Encoding by SSP}\label{Sec:SSP}
From previous section, the sensing matrix $A$ is defined in Kronecker format in (\ref{PF4}), which maps sparse images ${\bf C}$ into the measurement domain ${\bf Y}_{\Omega}$, i.e. $A:\mathbb{R}^N\mapsto\mathbb{R}^{|\Omega|}$. This mapping should be one-to-one in order to guarantee the unique recoverability of the original image from the sampled measurements \cite{CandesRombergTao1:2006, KashinTemlyakov:2007}. Seminal works are done in \cite{Jokar:2010, DuarteBaraniuk:2011, CaiafaCichocki:2012} for sparse recovery of the Kronecker matrices using the mutual coherency bounds. However, such bounds are very loose and limited to certain applications e.g. Gaussian matrices \cite{BrucksteinDonohoElad:2009} and do not apply to general cases. We take different approach to guarantee the sparse recovery of the Kronecker samplings. First we identify the relation between Spark \cite{DonohoElad:2003, BrucksteinDonohoElad:2009} and SSP \cite{KashinTemlyakov:2007} to grantees one-to-one mapping and then provide tractable bound for any arbitrary sensing matrices.

Let us define the class of $k$-sparse signals with $N$ coefficients ($N$-length vector) by $\sum_{k}=\{c\in\mathbb{R}^N:\|c\|_0\leq k\}$. The unique mapping of $k$-sparse signal $c$ under the projection of the sensing matrix $y=Ac$ is guaranteed if the nullspace of the sensing matrix, i.e. $\mathcal{N}(A)=\{\eta\in\mathbb{R}^N: A\eta=0\}$, contains vectors with more than $2k$-sparsity, i.e. $\sum_{2k}\nsubseteq\mathcal{N}(A)$ \cite{KashinTemlyakov:2007}. Furthermore, by the definition of the Spark in \cite{DonohoElad:2003}, i.e.
\begin{equation}\label{E1_SSP}
\text{Spark}(A) = \min_{\eta\in\mathcal{N}(A)}{\|\eta\|_0},
\end{equation}
the unique encoding is satisfied if $2k<\text{Spark}(A)$. The problem in (\ref{E1_SSP}) is a combinatorial minimization and is NP-hard. Tractable bound is proposed by means of SDP relaxation for the Spark in \cite{LeeBresler:2008}, which is computationally inefficient for dimensions greater than $N>30$ i.e. at most for $\sim 5\times 5$ image pixels. To overcome this problem, the lower-bound of the Spark is found by the Spherical Section Property (SSP) to provide computationally efficient methods for tracking unique recovery bounds. The SSP is defined as follows \cite{KashinTemlyakov:2007},
\begin{equation}\label{E2_SSP}
\text{SSP}(A) = \min_{\eta\in\mathcal{N}(A)\backslash\{0\}}{\frac{\|\eta\|_1}{\|\eta\|_2}}.
\end{equation}
If $\text{SSP}(A)\geq\Delta$, then $\Delta$ is a lower-bound and is known to be the distortion value of the nullspace $\mathcal{N}(A)$. High distortion values of $\Delta$ refers to almost an Euclidean nullspace, which contain vectors $\eta$ with high cardinalities. This implies that most sparse vectors lie on the range space of the sensing matrix rather than the nullspace. Therefore, higher the $\Delta$ is, better the sensing matrix $A$ is constructed for CS recovery.

The nullspace vector $\eta\in\mathcal{N}(A)$ can be extended as follows $\|\eta\|_1=\sum^{N}_{j=1}{|\eta_j|}= \sum^{N}_{j=1}{\eta_j\text{sign}(\eta_j)}$ and using the Cauchy-Shwartz inequality, it yields
\begin{equation}\label{E3_SSP}
\|\eta\|^2_1\leq\left[\sum^{N}_{j=1}{\text{sign}^2(\eta_j)}\right]^2
\left[\sum^{N}_{j=1}{\eta^2_j}\right]^2=\|\eta\|_0\|\eta\|^2_2.
\end{equation}
Substituting (\ref{E3_SSP}) in (\ref{E1_SSP}), the lower bound for Spark is provided by SSP i.e. $\Delta^2\leq\text{SSP}^2(A)\leq\text{Spark}(A)$. Following the above analysis, the sufficient condition for unique sparse encoding is carried out as follows.
\begin{proposition}
A system of linear equation $Ac=y$ has the unique sparse solution, iff
\begin{equation}\label{E4_SSP}
k<\frac{\Delta^2}{2}\leq\frac{\text{SSP}^2(A)}{2}
\end{equation}
where, $A\in\mathbb{R}^{|\Omega|\times{N}}$, $c\in\mathbb{R}^{N}$, $y\in\mathbb{R}^{|\Omega|}$ and ${|\Omega|<N}$.
\end{proposition}
The bound in (\ref{E4_SSP}) defines the maximum allowable sparsity rate of the solution $c$ to satisfy the unique sparse recovery, which can be at most $N/2$. Also, following the work in \cite{Jokar:2010}, the bound in (\ref{E4_SSP}) is extendable to Kronecker matrices, where $A=A_2\otimes A_1$. So, the sufficient condition for the linear system in (\ref{PF4}) to have the unique sparse solution is
\begin{equation}\label{E5_SSP}
k<\min_{i\in\{1,2\}}\{\Delta_i/2\}\leq \min_{i\in\{1,2\}}\{\text{SSP}^2(A_i)/2\},
\end{equation}
where, $\Delta_i$ refers to the lower bound of the $\text{SSP}(A_i)$ for mode-$i$ in 2D format. The lower bound $\Delta$ of the SSP can be quantized by the relaxed SDP problem, proposed in \cite{HosseiniFazeliPlataniotis:2012}, to study the recovery performance of the compressed sensing matrices. Seminal works exist in \cite{AspremontGhaoui:2011, LeeBresler:2008, TangNehorai:2011} for equivalent relaxed-SDP problems, where they have either complex SDP structure or practically inefficient for higher dimension. The proposed relaxed-SDP in \cite{HosseiniFazeliPlataniotis:2012} is driven by augmenting the LP and SDP cones together with $\mathcal{O}(N^2)$ and $\mathcal{O}(N^4)$ complexities, respectively. Hence, the problem is  large-scale and practically is capable of tracking the lower bound of the sensing matrix $A$ up to $N\sim 2000$ in ordinary desktop machines.

\section{Experimental Evaluations}\label{Sec:exp}
Our goal is to validate the recoverability of the different CS sampling masks, demonstrated in figure \ref{fig:Masks}, which can be applied in different practical scenarios e.g. CI cameras, medical imaging: MRI and tomography, and coded apertures. In particular, we evaluate the $\text{SSP}(A)$ lower bound, i.e. $\Delta$, for unique recovery in (\ref{E4_SSP}) by the proposed relaxed SDP problem in \cite{HosseiniFazeliPlataniotis:2012} for the aforementioned CS masks. For every constructed sensing matrix $A$, we also examine the sparse recovery by means of $\ell_1$-minimization in \cite{BergFriedlander:2008} using Basis Pursuit (BP) problem in order to validate the aforementioned lower bound $\Delta$. We define the size of the original image ${\bf X}\in\mathbb{R}^{32\times 32}$ consisting of $N=1024$ mesh-grids. We evaluate the SDP and $\ell_1$-minimization problems on different sampling ratios $|\Omega|/N=\{0.1,0.2,\hdots,0.6\}$ for $10$ randomly generated masks (except radial and down-sampling masks) and average the results. The results in Figure \ref{Fig:SSP_CI}-\ref{Fig:SSP_CP} demonstrate the normalized unique recovery bound in (\ref{E4_SSP})  i.e. $\Delta^2/2$ versus the sampling ratio. Also, results in Figure \ref{Fig:L1_CI}-\ref{Fig:L1_CA} demonstrate the sparse recovery using the BP problem.

\subsection{CI Camera}\label{SubSec:CI}
In Compressed Imaging (CI) camera setup, the goal is to reconstruct the image from its compressed measurements via randomly projected lights into a single/multiple sensors \cite{DuarteDavenportTakharLaskaTingKellyBaraniuk:2008, OikeGamal:2013}. The image is observed by programmable Digital Micromirror Device (DMD) with variety of possible transfer functions for sampling e.g. random i.i.d Gaussian/Bernoulli and Walsh-Hadamard bases \cite{DuarteDavenportTakharLaskaTingKellyBaraniuk:2008}. We assume $\Psi=W$ Daubechies$-4$ wavelet transform with $1$ level of sparse decomposition. Random CS mask for ${\bf s}^{\text{Grid}_{\Omega}}$ is taken from Figure \ref{Fig:BernoulliMask} for sampling. The lower bound $\Delta$ of the SSP for the sensing matrix $A$ in (\ref{PF4}) is evaluated and shown in Figure \ref{Fig:SSP_CI}. The results of BP problem also shown in Figure \ref{Fig:L1_CI}. The results from both relaxed-SDP and BP problems suggest that using Walsh-Hadamard bases for sampling is capable of recovering low sparsity rate of images (high cardinality) better than the other two bases: Gaussian/Bernoulli. The benefit of deploying Walsh-Hadamard is we don't need to restore the basis matrix and this can be effectively computed for high-dimensional image reconstructions in CS decodings.

\subsection{Rapid MRI}
In magnetic resonance imaging (MRI) body organs are scanned for clinical applications by collecting the images in Fourier domain (aka k-space) and, therefore, the corresponding transfer function $\phi$ is derived from the Fourier basis kernels \cite{LustigDonohoPauly:2007}. Here, we consider $\Phi=\mathcal{F}$, which is $32\times 32$ discrete Fourier basis matrix. for this experiment we use the same wavelets for sparse decomposition in Section \ref{SubSec:CI}. we examine three CS masks for under-sampling: radial (maximum of 22 lines), random and density-varied random masks exemplified in Figure \ref{fig:Masks}. Different number of lines $\{2,4,\hdots,22\}$ are used for radial mask and the sampling ratio is defines by $|\Omega|/N$, where $\Omega$ corresponds to the indices occupied by the radial lines in the mask. The results of approximated lower bound $\Delta$ and $BP$ problems are shown in Figure \ref{Fig:SSP_MRI} and \ref{Fig:L1_MRI}, respectively. Results show that random mask with i.i.d Bernoulli distribution is slightly performing better than radial and density-varied masks. While designing the sampling mask in MRI can be complicated, our results suggest to perform sampling using random masks which takes samples on the periphery of the k-space.

\subsection{Coded Aperture}
\begin{figure}[h]
\centerline{
\subfigure[]{\includegraphics[width=0.152\textwidth]{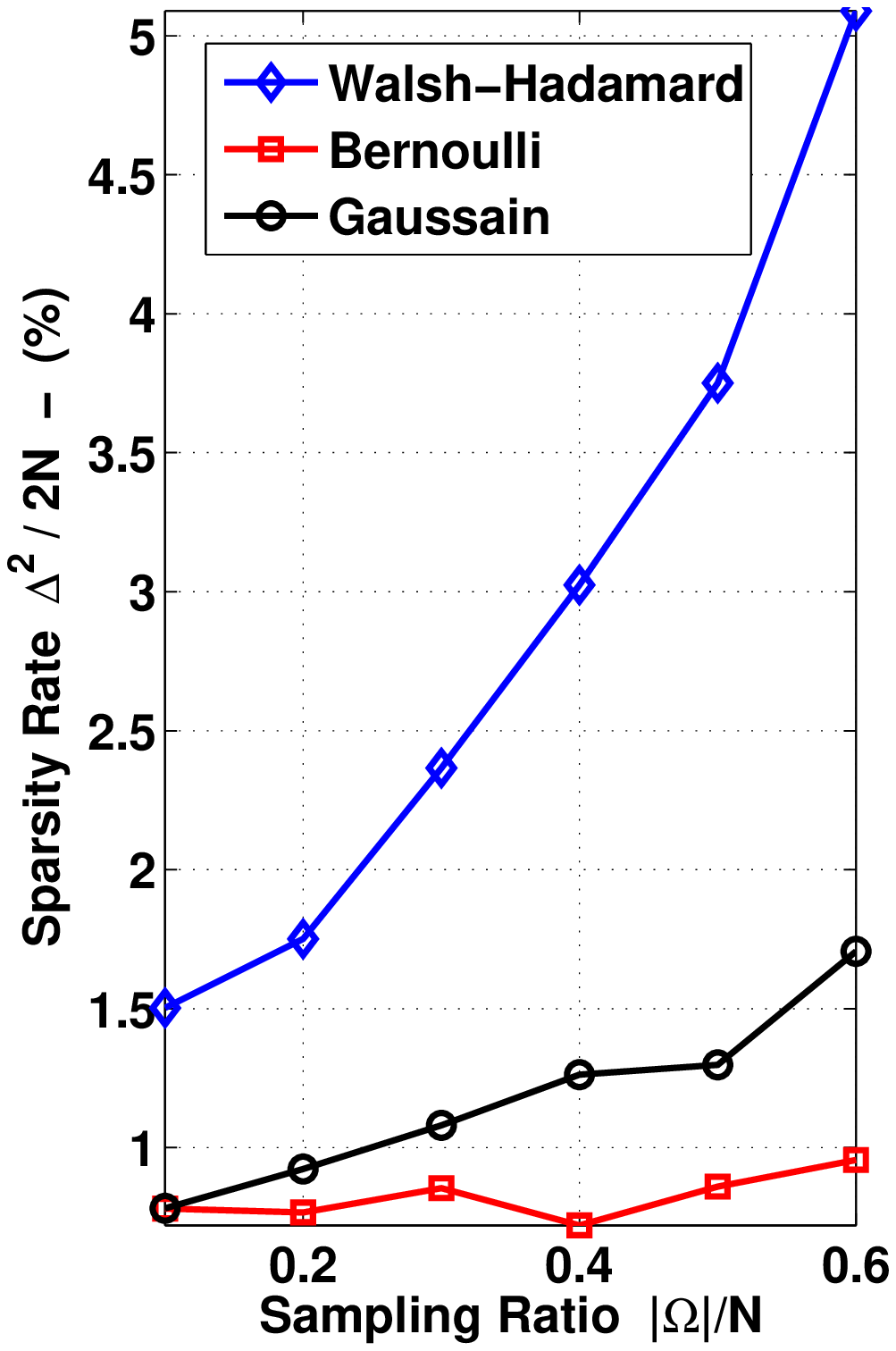}\label{Fig:SSP_CI}}
\subfigure[]{\includegraphics[width=0.152\textwidth]{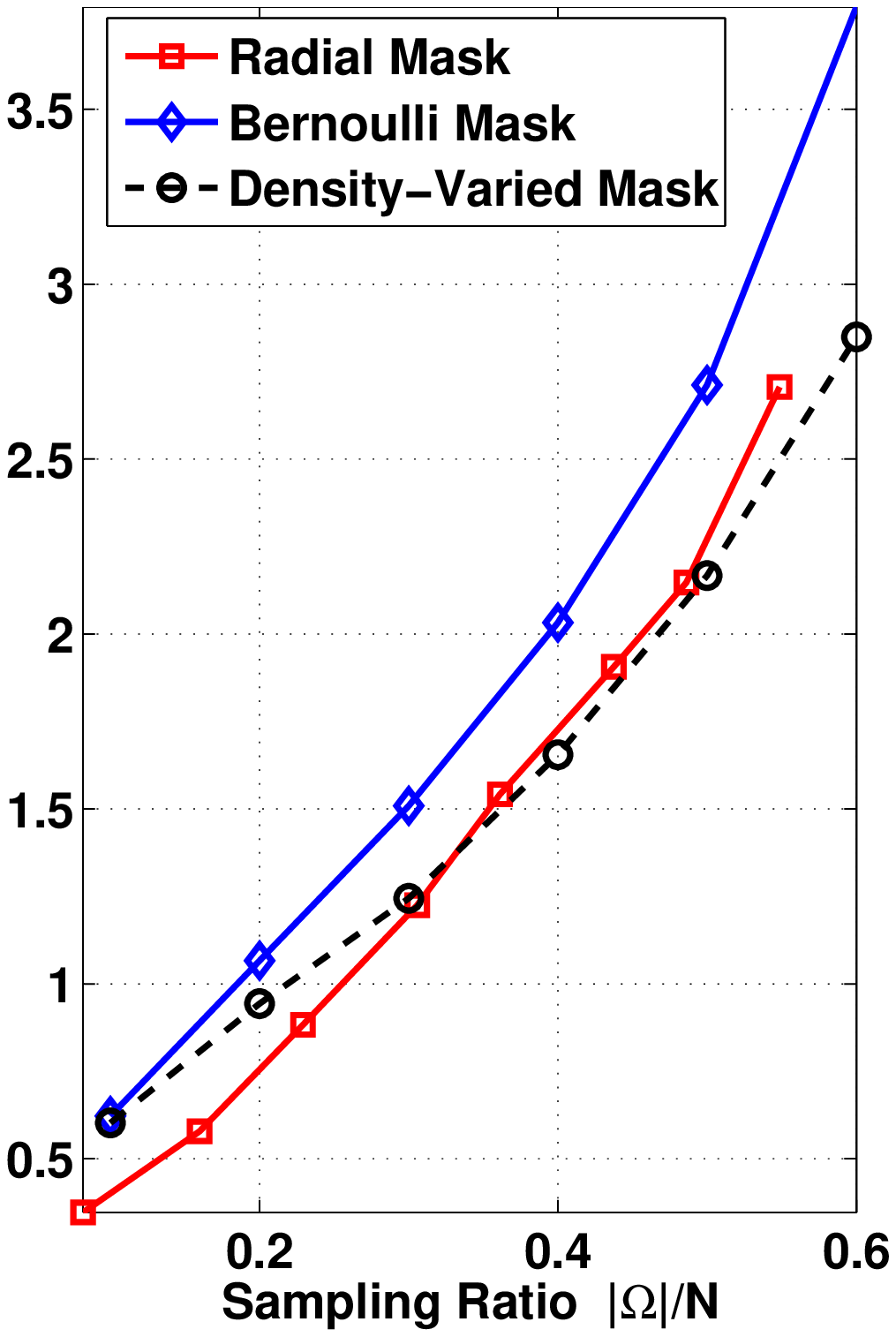}\label{Fig:SSP_MRI}}
\subfigure[]{\includegraphics[width=0.152\textwidth]{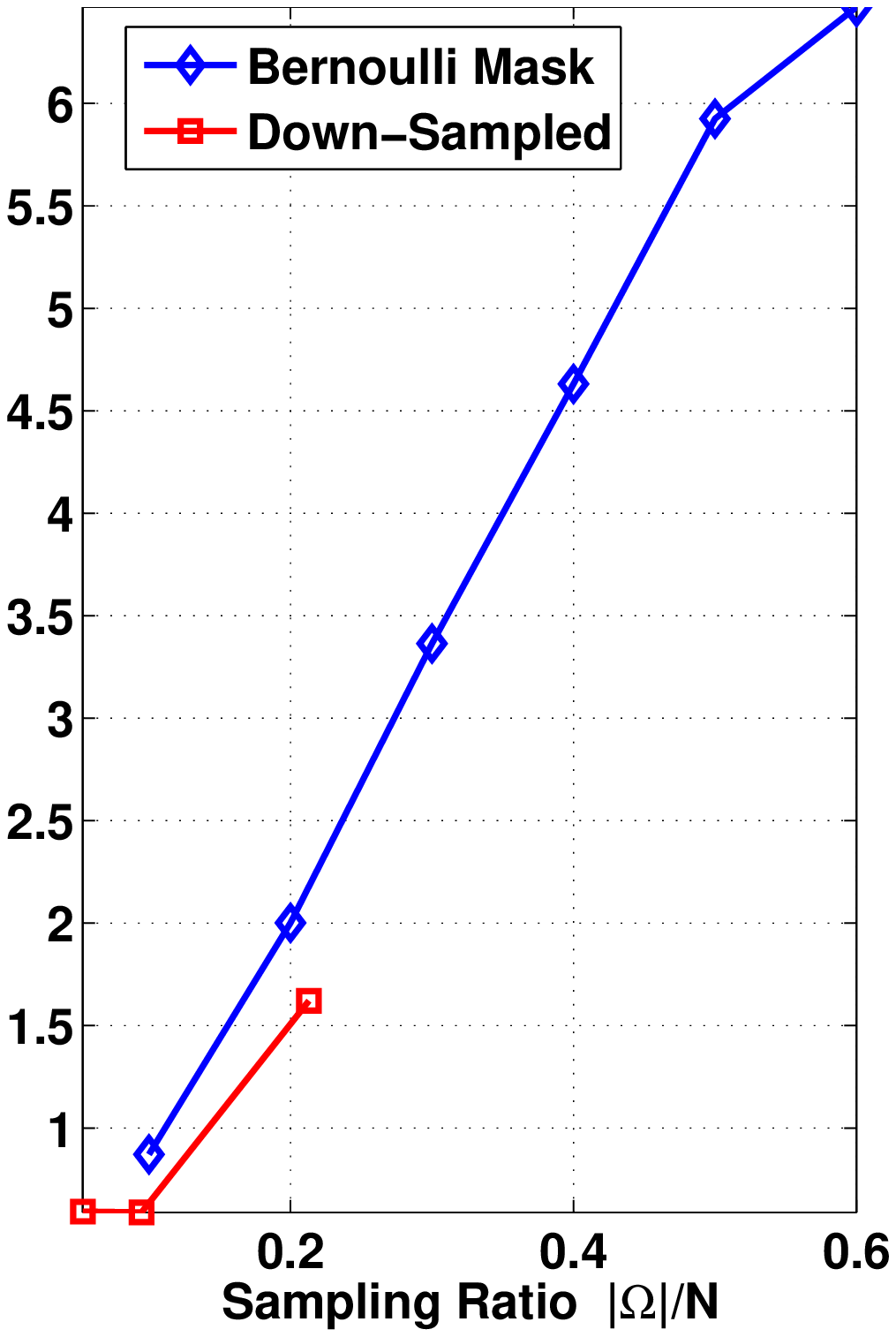}\label{Fig:SSP_CP}}
}
\centerline{
\subfigure[]{\includegraphics[width=0.152\textwidth]{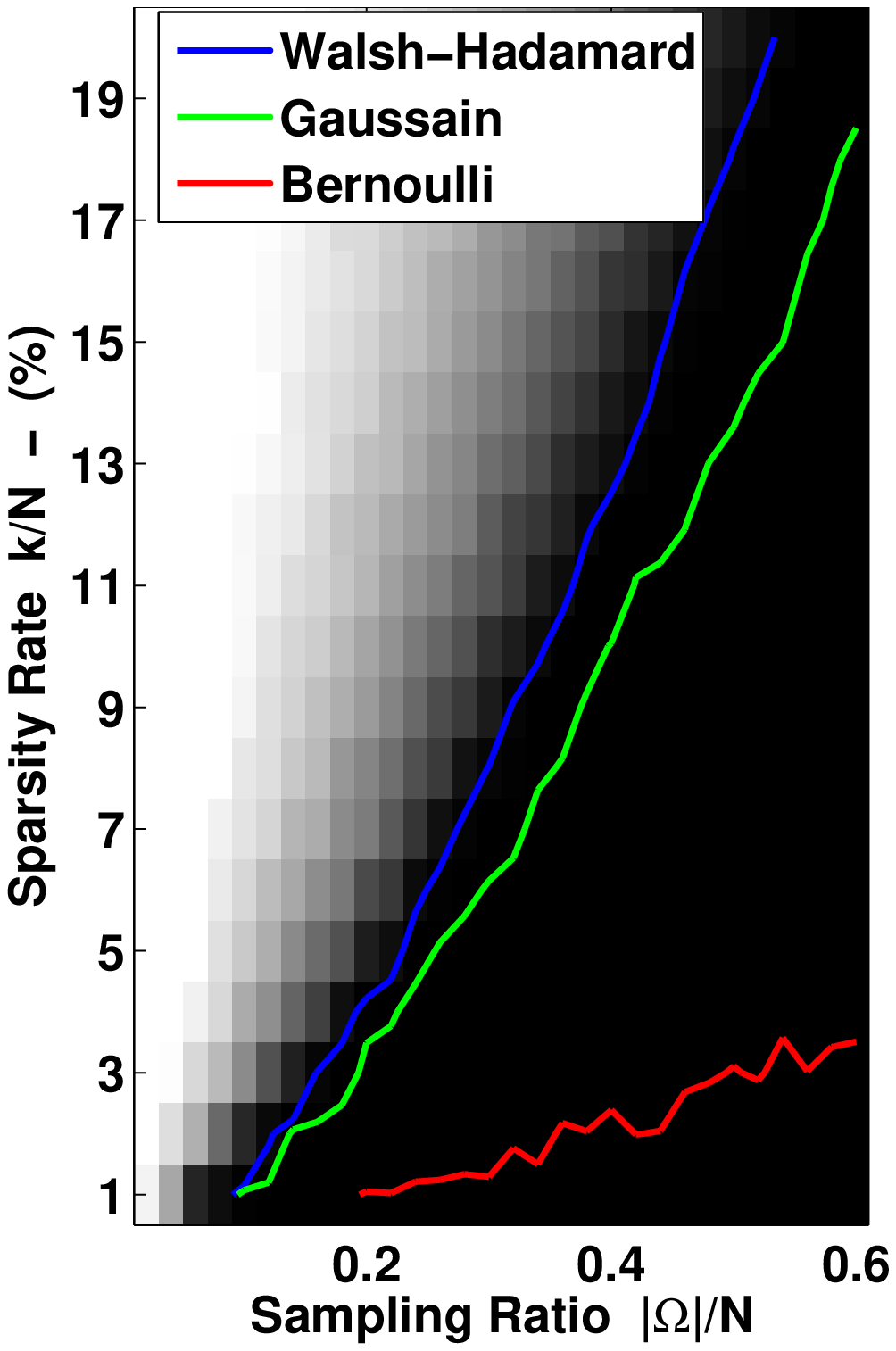}\label{Fig:L1_CI}}
\subfigure[]{\includegraphics[width=0.152\textwidth]{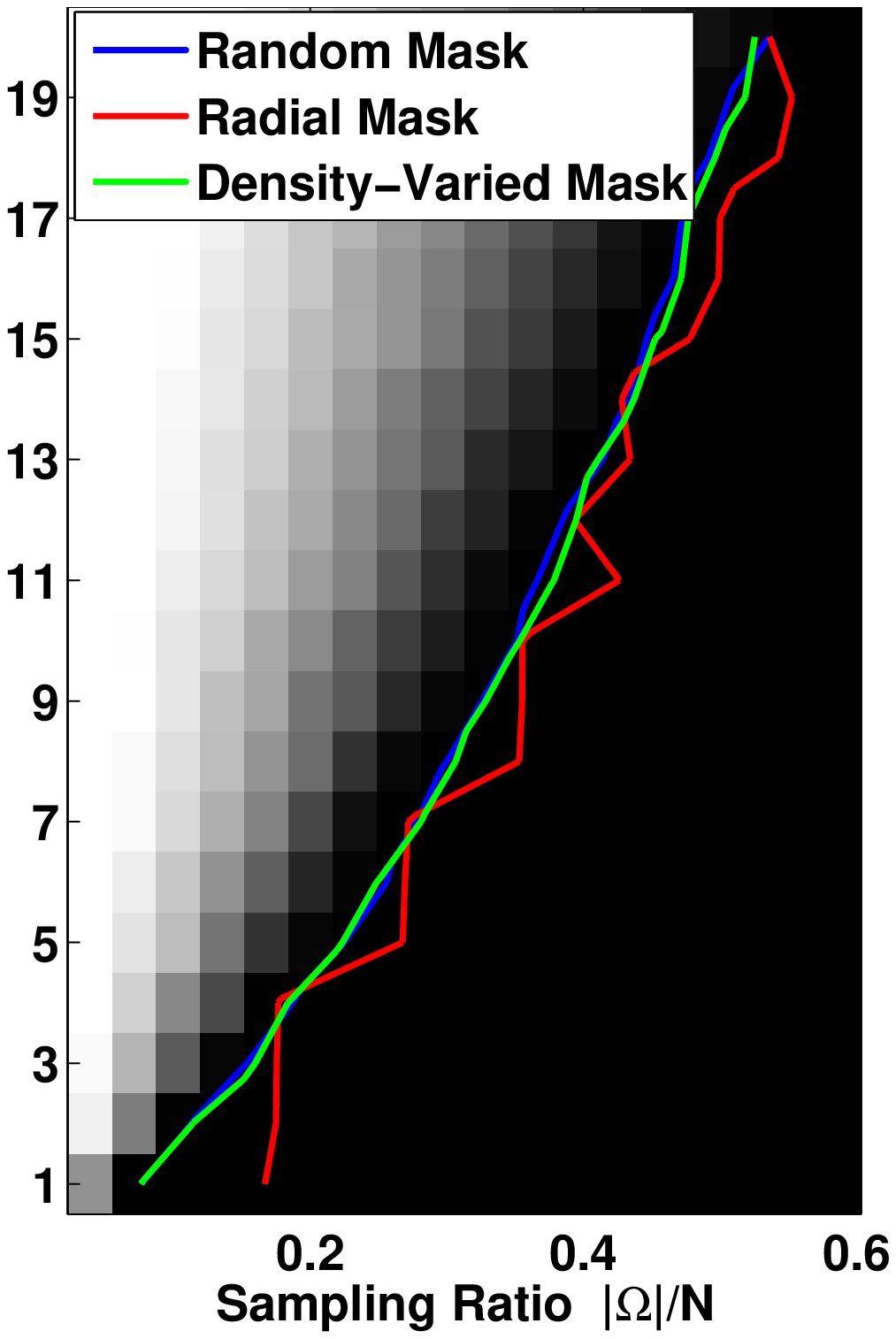}\label{Fig:L1_MRI}}
\subfigure[]{\includegraphics[width=0.152\textwidth]{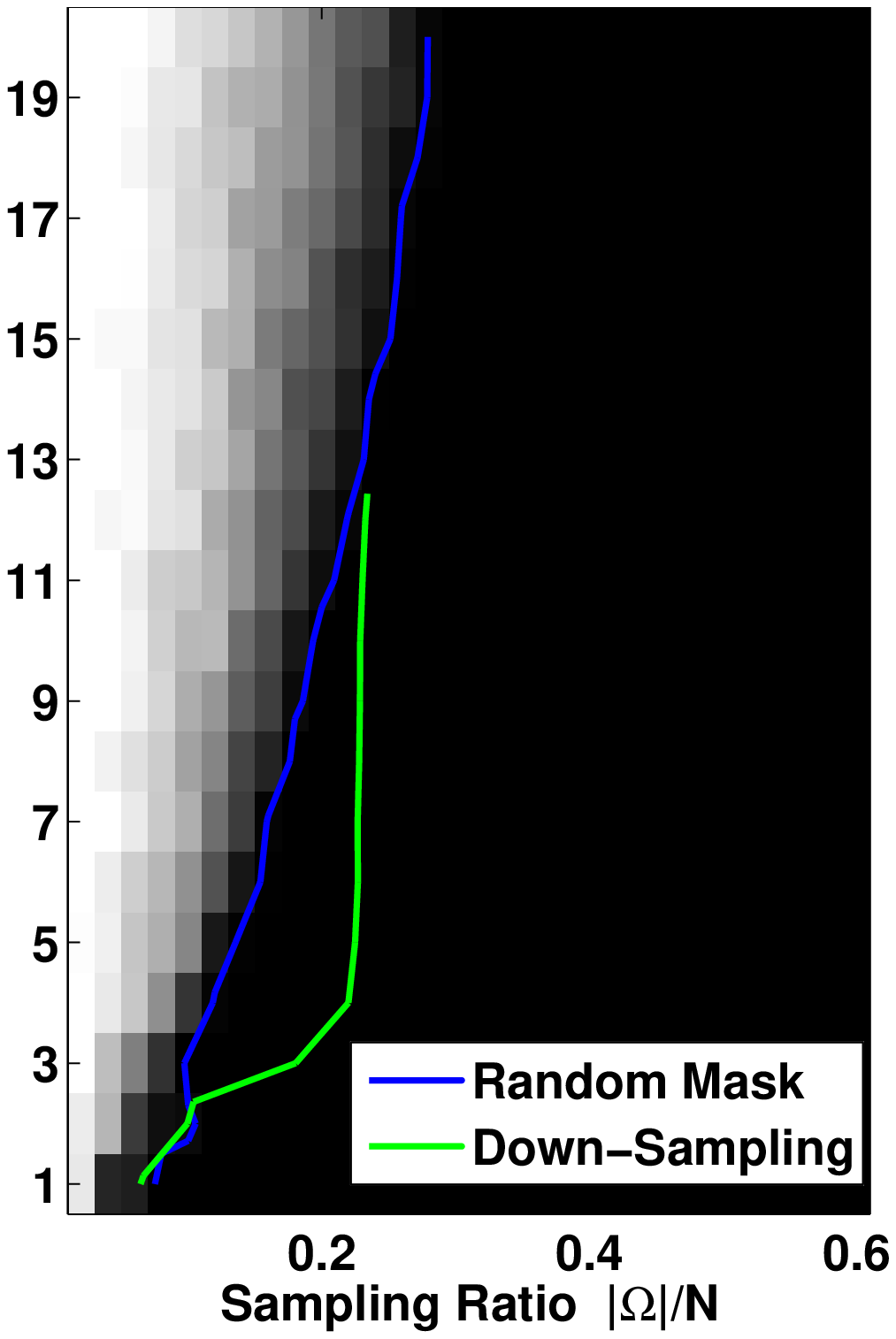}\label{Fig:L1_CA}}
}\vspace{-.1in}
\caption{Lower bound $\Delta$ by relaxed-SDP (first row) and phase transition of sparse recovery using $\ell_1$-minimization (second row) for: (a\&d) CI camera; (b\&e) MRI; (c\&f) Coded aperture.}
\label{Fig:SSP}
\end{figure}

The idea is to recover a high-resolution image from the low-resolution version. Due to the aperture problem the registered image is subjected to the law of diffraction in the camera. Such diffraction is modelled by a Point Spread Function (PSF), which is also know as the transfer function in optics and is modelled by a complicated coded mask e.g. MURA pattern in Figure \ref{fig:MURA}. The general framework for down-sampled (compressed) coded aperture frame is give by ${\bf Y}_\Omega=D({\bf X}\ast \text{PSF})$, where $D$ is the down-sampling operator followed by the CS mask in Figure\ref{Fig:ApertureMask} \cite{MarciaWillett:2008}. The mask pattern should be prime-integer side length and hence we consider $m=n=31$ in our experiment. The objective of MURA pattern mask is to filter certain wavelength which is of interest in both astronomical and medical applications. The corresponding transfer function operator is defined by $\Phi=\mathcal{F}^{-1}C_{\text{PSF}}\mathcal{F}$, where $C_{\text{PSF}}$ is the MURA pattern. Results of the lower bound $\Delta$ are sketched in Figure \ref{Fig:SSP_CP} and the BP in Figure \ref{Fig:L1_CA}. The most sampling ratio from down-sampling is $25\%$ and we cannot go beyond than this ratio. We, also examine additional CS masks in Figure \ref{Fig:BernoulliMask} for random sampling to compare the results with downsampling method. The benefit of deploying this mask is to reach to higher performances by taking more samples, however, taking random samples requires a design of hardware setup in practice.

\vspace{-.15in}

\section{Concluding Remarks}
We have studied the unique recovery conditions of three existing CS imaging applications i.e. CI cameras, rapid MRI and coded apertures, by means of tracking the lower bound of the SSP of the sensing matrices using a relaxed-SDP and then validated through numerical experiments using Basis Pursuit (BP) problem for sparse image recovery. Walsh-Hadamard sampling, in CI camera, was capable of high order recoveries. This basis can be efficiently implemented in practice for decoding, since, there is no need to buffer the basis matrix and can be deployed as vector-matrix calculation. In MRI application, the performance of three sub-sampling pattern masks are studied and their effectiveness are verified. We noticed random mask performs slightly better than two other patterns i.e density-varied and radial. In coded aperture, the higher gain of recovery was observed using a circulant matrix with PSF kernel. We suggest CS practitioners to consider this sampling architecture for further investigations in practical problems.

\bibliographystyle{IEEEbib}
\bibliography{myref}

\begin{thebibliography}{10}

\bibitem{CandesTao:2005}
E.J. Candes and T.~Tao,
\newblock ``Decoding by linear programming,''
\newblock {\em Information Theory, IEEE Transactions on}, vol. 51, no. 12, pp.
  4203 -- 4215, December 2005.

\bibitem{KashinTemlyakov:2007}
B.~Kashin and V.~Temlyakov,
\newblock ``A remark on compressed sensing,''
\newblock {\em Mathematical Notes}, vol. 82, pp. 748--755, 2007.

\bibitem{Zhang:2008}
Yin Zhang,
\newblock ``Theory of compressive sensing via l1-minimization: A non-rip
  analysis and extensions,''
\newblock {\em Technical report, Rice University}, 2008.

\bibitem{BrucksteinDonohoElad:2009}
Alfred~M. Bruckstein, David~L. Donoho, and Michael Elad,
\newblock ``From sparse solutions of systems of equations to sparse modelling
  of signals and images,''
\newblock vol. 51, no. 1, pp. 34--81, 2009.

\bibitem{CandesRombergTao1:2006}
E.J. Candes, J.~Romberg, and T.~Tao,
\newblock ``Robust uncertainty principles: exact signal reconstruction from
  highly incomplete frequency information,''
\newblock {\em Information Theory, IEEE Transactions on}, vol. 52, no. 2, pp.
  489 -- 509, February 2006.

\bibitem{DuarteEldar:2011}
M.F. Duarte and Y.C. Eldar,
\newblock ``Structured compressed sensing: From theory to applications,''
\newblock {\em Signal Processing, IEEE Transactions on}, vol. 59, no. 9, pp.
  4053 --4085, September 2011.

\bibitem{DuarteDavenportTakharLaskaTingKellyBaraniuk:2008}
M.F. Duarte, M.A. Davenport, D.~Takhar, J.N. Laska, Ting Sun, K.F. Kelly, and
  R.G. Baraniuk,
\newblock ``Single-pixel imaging via compressive sampling,''
\newblock {\em Signal Processing Magazine, IEEE}, vol. 25, no. 2, pp. 83 --91,
  March 2008.

\bibitem{OikeGamal:2013}
Y.~Oike and A.~El~Gamal,
\newblock ``Cmos image sensor with per-column sigma-delta; adc and programmable
  compressed sensing,''
\newblock {\em Solid-State Circuits, IEEE Journal of}, , no. 99, pp. 1 --11,
  January 2013.

\bibitem{LustigDonohoPauly:2007}
Michael Lustig, David Donoho, and John~M. Pauly,
\newblock ``Sparse mri: The application of compressed sensing for rapid mr
  imaging,''
\newblock {\em Magnetic Resonance in Medicine}, vol. 58, no. 6, pp. 1182--1195,
  2007.

\bibitem{MarciaWillett:2008}
R.F. Marcia and R.M. Willett,
\newblock ``Compressive coded aperture superresolution image reconstruction,''
\newblock in {\em Acoustics, Speech and Signal Processing, 2008. ICASSP 2008.
  IEEE International Conference on}, April 2008, pp. 833 --836.

\bibitem{Jokar:2010}
S.~Jokar,
\newblock ``Sparse recovery and kronecker products,''
\newblock in {\em Information Sciences and Systems (CISS), 2010 44th Annual
  Conference on}, March 2010, pp. 1--4.

\bibitem{DuarteBaraniuk:2011}
M.~Duarte and R.~Baraniuk,
\newblock ``Kronecker compressive sensing,''
\newblock {\em Image Processing, IEEE Transactions on}, vol. PP, no. 99, pp. 1,
  2011.

\bibitem{CaiafaCichocki:2012}
Cesar~F. Caiafa and Andrzej Cichocki,
\newblock ``Computing sparse representations of multidimensional signals using
  kronecker bases,''
\newblock {\em Neural Computation}, vol. 1, no. 35, 2012.

\bibitem{DonohoElad:2003}
David~L. Donoho and Michael Elad,
\newblock ``Optimally sparse representation in general (nonorthogonal)
  dictionaries via l1 minimization,''
\newblock {\em Proceedings of the National Academy of Sciences of the United
  States of America}, vol. 100, no. 5, pp. pp. 2197--2202, 2003.

\bibitem{AspremontGhaoui:2011}
Alexandre d'Aspremont and Laurent El~Ghaoui,
\newblock ``Testing the nullspace property using semidefinite programming,''
\newblock {\em Mathematical Programming}, vol. 127, pp. 123--144, 2011.

\bibitem{Aspremont:2011}
A.~{d'Aspremont},
\newblock ``Sparse recovery, kashin decomposition and conic programming,''
\newblock {\em ArXiv e-prints}, January 2011.

\bibitem{HosseiniFazeliPlataniotis:2012}
M.~S. Hosseini, S.~Fazeli-Dehkordy, and K.~N. Plataniotis,
\newblock ``Tractable bound for spherical section property in the presence of
  side-information,''
\newblock {\em Signal Processing Letters, IEEE}, vol. 19, no. 8, pp. 519 --522,
  August 2012.

\bibitem{LeeBresler:2008}
Kiryung Lee and Y.~Bresler,
\newblock ``Computing performance guarantees for compressed sensing,''
\newblock in {\em Acoustics, Speech and Signal Processing, 2008. ICASSP 2008.
  IEEE International Conference on}, April 2008, pp. 5129 --5132.

\bibitem{TangNehorai:2011}
Gongguo Tang and A.~Nehorai,
\newblock ``Performance analysis of sparse recovery based on constrained
  minimal singular values,''
\newblock {\em Signal Processing, IEEE Transactions on}, vol. 59, no. 12, pp.
  5734 --5745, December 2011.

\bibitem{BergFriedlander:2008}
Ewout van~den Berg and Michael~P. Friedlander,
\newblock ``Probing the pareto frontier for basis pursuit solutions,''
\newblock vol. 31, no. 2, pp. 890--912, 2008.

\bibitem{HahnLimChoiHorisakiBrady:2011}
Joonku Hahn, Sehoon Lim, Kerkil Choi, Ryoichi Horisaki, and David~J. Brady,
\newblock ``Video-rate compressive holographic microscopic tomography,''
\newblock {\em Opt. Express}, vol. 19, no. 8, pp. 7289--7298, April 2011.

\bibitem{ReddyVeeraraghavanChellappa:2011}
D.~Reddy, A.~Veeraraghavan, and R.~Chellappa,
\newblock ``P2c2: Programmable pixel compressive camera for high speed
  imaging,''
\newblock in {\em Computer Vision and Pattern Recognition (CVPR), 2011 IEEE
  Conference on}, June 2011, pp. 329 --336.

\bibitem{BourquardAguetUnser:2010}
Aur\'{e}lien Bourquard, Fran\c{c}ois Aguet, and Michael Unser,
\newblock ``Optical imaging using binary sensors,''
\newblock {\em Opt. Express}, vol. 18, no. 5, pp. 4876--4888, March 2010.

\bibitem{Brewer:1978}
J.~Brewer,
\newblock ``Kronecker products and matrix calculus in system theory,''
\newblock {\em Circuits and Systems, IEEE Transactions on}, vol. 25, no. 9, pp.
  772 -- 781, September 1978.

\end{thebibliography}

\end{document}